\documentclass[aps,onecolumn,groupedaddress,nofootinbib,amsmath,amssymb]{revtex4}
\usepackage{dsfont}
\usepackage{bm}
\begin{document}

\title{Deeper discussion  of Schr\"odinger invariant and Logarithmic sectors of
higher-curvature gravity}

\author{Eloy Ay\'on--Beato}\email{ayon-beato-at-fis.cinvestav.mx}
\affiliation{Departamento~de~F\'{\i}sica,~CINVESTAV--IPN,~
Apdo.~Postal~14--740,~07000,~M\'exico~D.F.,~M\'exico.}

\author{Gaston Giribet}\email{gaston-at-df.uba.ar}
\affiliation{Instituto de F\'{\i}sica de Buenos Aires IFIBA -
CONICET, Buenos Aires, Argentina.}

\author{Mokhtar Hassa\"{\i}ne}\email{hassaine-at-inst-mat.utalca.cl}
\affiliation{Instituto de Matem\'atica y F\'{\i}sica, Universidad de
Talca, Casilla 747,
Talca, Chile.\\
Laboratoire de Math\'ematiques et de Physique Th\'eorique,
Universit\'e de Tours, Parc de Grandmont, 37200 Tours, France.}

\begin{abstract}
The aim of this paper is to explore $D$-dimensional theories of pure
gravity whose space of solutions contains certain class of
AdS-waves, including in particular Schr\"odinger invariant
spacetimes. This amounts to consider higher order theories, and the
natural case to start with is to analyze generic square-curvature
corrections to Einstein-Hilbert action. In this case, the
Schr\"odinger invariant sector in the space of solutions arises for
a special relation between the coupling constants appearing in the
action. On the other hand, besides the Schr\"odinger invariant
configurations, logarithmic branches similar to those of the
so-called Log-gravity are also shown to emerge for another special
choice of the coupling constants. Interestingly enough, these Log
solutions can be interpreted as the superposition of the massless
mode of General Relativity and two scalar modes that saturate the
Breitenlohner-Freedman bound (BF) of the AdS space on which they
propagate. These solutions are higher-dimensional analogues of those
appearing in three-dimensional massive gravities with relaxed
AdS$_3$ asymptotic, which are candidates to be gravity duals for
logarithmic CFTs. Other sectors of the space of solutions of
higher-curvature theories correspond to oscillatory configurations,
which happen to be below the BF bound. Also, there is a fully
degenerated sector, for which any wave profile is admitted. We
comment on the relation between this degeneracy and the
non-renormalizability of the dynamical exponent of the Schr\"odinger
spaces. Our analysis also includes more general gravitational
actions with non-polynomial corrections consisting of arbitrary
functions of the square-curvature invariants. By establishing a
correspondence with the quadratic gravity model, the same sectors of
solutions are shown to exist for this more general family of
theories. We finally consider the parity-violating Chern-Simons
modified gravity in four dimensions, for which we derive both the
Schr\"odinger invariant as well as the logarithmic sectors.

{\it This paper is dedicated to the memory of Laurent Houart.}
\end{abstract}

\maketitle

\section{Introduction}

The AdS/CFT correspondence
\cite{Maldacena,Gubser,Witten,Aharony:1999ti} provides a dictionary
to relate conformal field theories in flat space to
higher-dimensional gravitational theories. In this framework, the
gravitational description is weakly coupled when the relativistic
CFT is strongly coupled, and thus it becomes a promising tool to
explore fundamental physics in the non-perturbative regime. However,
in spite of the enormous mathematical success of AdS/CFT
correspondence, the experimental applications of this idea have
somewhat been braked by the fact that only few relativistic
conformal field theories at strong coupling are accessible
experimentally.

Nevertheless, unlike what happens in the relativistic case, there
exists a plenitude of non-relativistic conformal field models that
govern physics in different experimentally accessible areas such as
condensed matter physics and atomic or nuclear physics. From this
perspective, the idea of generalizing AdS/CFT correspondence to the
case of non-relativistic conformal field theories has been proposed
\cite{Son:2008ye,BMC}. In the non-relativistic case, the symmetry
group is identified with the conformal extension of the Galilei
group, often called the Schr\"odinger group. This identification is
due to the fact that there exist some analogies between the
Minkowski conformal algebra and the Schr\"odinger algebra in one
dimension less. This analogy between both algebras has motivated the
search of a geometric framework to understand the Schr\"odinger
algebra. One pioneer work in this direction was the one done by
Havas and Pleba\'nski \cite{HP}, where they proposed to introduce
the Schr\"odinger symmetry as a sub-group of the
infinite-dimensional group of Galilean conformal transformations of
flat spacetime. A geometric picture of the Schr\"odinger algebra and
its relations with the conformal algebra have been explained in
\cite{Duval:1990hj}, and this result was used in \cite{Duval:2008jg}
to derive the invariant Schr\"odinger metric in the context of the
non-relativistic AdS/CFT correspondence. In turn, following the
philosophy of the AdS/CFT holographic correspondence, the hope is
that backgrounds whose (asymptotic) isometry group agrees with the
Schr\"odinger group would represent the gravity duals of some
conformal quantum mechanical systems with applications to condensed
matter physics.\footnote{See \cite{Guica} and \cite{McGreevyNew} for
interesting discussions on holography in Schr\"odinger spaces.}
Owing to this expectation, there has been an increasing interest in
constructing solutions of string theory inspired models whose
(asymptotic boundary) isometry group is given by the Schr\"odinger
group. For instance, the embedding of such solutions and their
analogues at finite temperature in string theory were considered in
\cite{Herzog:2008wg,Maldacena:2008wh,Adams:2008wt} (see also
references thereof). The effects of string inspired higher-curvature
corrections of these spaces were first analyzed in
\cite{Adams:2008zk}, where it was shown that such higher order
corrections lead to renormalize the dynamical exponent of the dual
conformal field theory. Here, we go deeper into the discussion on
higher-curvature actions and we argue that Schr\"odinger invariant
backgrounds arise as solutions of different theories of pure gravity
in four and higher dimensions. It is relatively easy to show that
Schr\"odinger invariant spacetimes arise as solutions of Einstein
gravity with negative cosmological constant, provided the support of
matter fields. This can be achieved with the introduction of some
reasonable physical source, like a Proca field \cite{Son:2008ye}
and/or an Abelian Higgs field in its broken phase \cite{BMC}.
However, solutions enjoying Schr\"odinger symmetry are also possible
in theories of gravity in absence of matter. The simplest example of
a {\it pure} gravity theory for which solutions possessing full
Schr\"odinger symmetry have been found is three--dimensional
Topologically Massive Gravity (TMG) with negative cosmological
constant \cite{Deser:1981wh}. This theory of gravity has attracted
considerable attention the last three years, and, in particular, one
of the interesting features of TMG is precisely that it admits exact
AdS wave solutions
\cite{AyonBeato:2004fq,Dereli:2000fm,Olmez:2005by,AyonBeato:2005qq,%
Carlip:2008eq,Gibbons:2008vi}; see below for a precise definition of
such spacetimes. While for generic values of the topological mass
$\mu$ and the cosmological constant $\Lambda = -1/l^2$ these AdS
wave solutions are only \emph{partially\/} Schr\"odinger invariant
\footnote{As shown below, this corresponds to the Galilei
transformations together with an anisotropic rescaling}, it turns
out that for the special fine tuning $\mu l=3$ the solution exhibits
the \emph{full\/} Schr\"odinger symmetry. This fine tuning
corresponds to the critical point of the space of parameters at
which the warped AdS$_3$ solution of the theory has isometry group
$\mathop{\rm SL}(2,\mathbb{R})\times\mathop{\rm U}(1)$ with a null
$\mathop{\rm U}(1)$ direction \cite{Anninos:2008fx}. More recently,
in Ref.~\cite{AyonBeato:2009yq}, the authors of the present paper
have shown the existence of Schr\"odinger invariant spaces for the
so-called New Massive Gravity (NMG) introduced in
Ref.~\cite{Bergshoeff:2009hq}.

Nevertheless, while the Schr\"odinger invariant solutions of TMG and
NMG are interesting in their own right, their relevance concerning
the non-relativistic holographic correspondence one is trying to
construct is somewhat questionable. This is basically because its
non-relativistic dual model would correspond to a theory with zero
spatial dimensions. In turn, it is natural to ask for a
higher-dimensional extension of this construction: Is there a theory
of pure gravity in four (or higher) dimensions that admits solutions
with Schr\"odinger isometry group? Actually, one can answer in the
affirmative. One of the ideas of this paper is to look for such
theories and go deepen into the Schr\"odinger invariant sector of
pure gravity.

We will argue that it is actually large the class of theories of
higher-curvature gravity that admits solutions with Schr\"odinger
isometry group in four and higher dimensions. To construct a theory
of pure gravity that admits exact solutions with Schr\"odinger
isometry group in $D>3$ dimensions, one can follow two different
strategies: The first one is resorting to the results of
Ref.~\cite{Adams:2008zk}. There, it was shown that the inclusion of
higher-curvature terms in the gravitational action leads to
renormalize the dynamical exponent of the Schr\"odinger symmetric
spaces (this exponent is usually denoted by $z=\nu+1$). Then, one
could in principle make inverse engineering and use the running
equation that relates the coupling constants of the gravitational
Lagrangian to the dynamical exponent $\nu$ in order to design the
model. Here, instead, we will follow a rather different strategy. We
will scan a wider class of higher-order gravitational Lagrangians
and exhaustively analyze the sectors of solutions that belong to a
special type of Siklos spacetimes. This will lead us to find several
sectors of the space of solutions including, in particular,
Schr\"odinger spaces, together with logarithmic solutions similar to
those found in three dimensions.

The study of the higher-dimensional case starts in
Sec.~\ref{sec:D>4} with an analysis of the most general modified
theory of gravity with quadratic dependence on the curvature. We
find the explicit configurations for generic dimension, which allow
us to identify several different sectors, apart from the
Schr\"odinger invariant one. For example, there are critical points
in the space of coupling constants for which a full degeneracy
arises and the field equations are not only satisfied for any
dynamical exponent $z$ but even for any AdS-wave profile function
$F$. The corresponding set of theories includes (but is not limited
to) the usual degenerate case of Chern-Simons gravity in $D=5$. In
analogy with the three-dimensional configurations of TMG
\cite{AyonBeato:2004fq,AyonBeato:2005qq} and NMG
\cite{AyonBeato:2009yq},\footnote{See
Refs.~\cite{Afshar:2009rg,Alishahiha:2010iq} for similar
configurations appearing in bi-gravity and Born-Infeld gravity
\cite{Gullu:2010pc}.} we observe also the existence of other
critical values of the coupling constants for which Schr\"odinger
invariance is broken by the appearance of logarithmic behaviors
associated to the existence of scalar wave modes that saturate the
Breitenlohner-Freedman bound (BF) of the AdS$_D$ space on which the
wave propagates. Some of these logarithmic decays are candidate to
relax the usual AdS boundary conditions
\cite{Henneaux:1985tv,Brown:1986nw,Henneaux:1985ey} in the context
of theories with higher-curvature corrections, in perfect analogy
with what occurs in three-dimensional massive gravities
\cite{Grumiller:2008qz,Henneaux:2009pw}.

In Sec.~\ref{sec:mggt}, we further generalize the results of section
3 by studying non-polynomials gravity modifications which depend
arbitrarily on the squared-curvature invariants; this is done by
establishing a correspondence between such a theory and the one with
standard quadratic modifications. Since Siklos spacetimes are
conformally equivalent to pp-waves, the metric configurations turn
out to be persistent when a wide class of higher-curvature
corrections are included, provided a suitable redefinition of the
parameters. Within this context, and even if the existence of
Schr\"odinger invariant sectors is not specially related to
conformal invariance, we find illustrative to explore in
Sec.~\ref{sec:Conf_corr} the particular example of adding to the
$D$-dimensional Einstein-Hilbert action the conformally invariant
deformation proportional to $(C_{\mu \nu \alpha \beta}C^{\mu \nu
\alpha \beta})^{D/4}$, where $C_{\mu \nu \alpha \beta}$ is the Weyl
tensor. Interestingly enough, we observe that only in four
dimensions the Schr\"odinger invariant sector, the logarithmic
sector, and the sector below the BF bound are genuinely new
configurations, while in arbitrary $D>4$ the resulting solutions
turn out to coincide with those of General Relativity. We will also
consider a rather different model of higher-curvature gravity in
Sec.~\ref{sec:C-Smod}, given by the parity-violating Chern-Simons
modification of General Relativity proposed by Jackiw and Pi
\cite{Jackiw:2003pm}, and we will show that Schr\"odinger invariant
and Log configurations are also admitted as solutions in this
theory. Finally, Section \ref{sec:conclu} is devoted to our
conclusions and further prospect. We include an Appendix describing
the behavior of higher-order curvature terms in presence of AdS
waves.

\textbf{Note added}. When our paper was being prepared for
publication, Refs.~\cite{Alishahiha:2011yb} and \cite{Gullu:2011sj}
appeared, which consider some of the configurations studied in this
work, including the relevant ones for the recently proposed critical
gravity theories \cite{Lu:2011zk,Deser:2011xc}.

\section{Schr\"odinger isometry group and the Siklos spacetimes
\label{sec:SSiklos}}

For the presentation to be self-contained, let us begin by reviewing
some aspects of the Schr\"odinger group. The Schr\"odinger group has
been defined in \cite{JackiwPT,Niederer:1972zz,Hagen:1972pd} as the
largest group of space-time transformations which leaves the
Schr\"odinger equation for a free particle invariant. Schr\"odinger
invariance has been considered in a wide variety of situations,
including celestial mechanics \cite{Duval:1990hj}, non-relativistic
field theory \cite{J,Jackiw:1990mb}, non-relativistic quantum
mechanics \cite{Ghosh:2001an}, and hydrodynamics
\cite{Hassaine:1999hn,O'Raifeartaigh:2000mp,Hassaine:2000ti,%
Nyawelo:2003bv,Montigny:2003az}. Mathematical aspects of the
Schr\"odinger symmetry have been analyzed, for instance, in
\cite{Henkel:2003pu}. The Schr\"odinger group can be viewed as the
semi-direct product of the connected static Galilei group together
with the $\mathop{\rm SL}(2,\mathbb{R})$ group, which includes time
translation, dilatation, and special conformal transformations. This
is an extension of the Lifshitz group, which is also considered in
the holographic description of non-relativistic models
\cite{Kachru:2008yh}.

As mentioned, the main idea of the proposal for a non-relativistic
version of the AdS/CFT correspondence \cite{Son:2008ye,BMC} is to
consider a metric whose isometry is given by the non-relativistic
conformal Schr\"odinger symmetry. In this context, we will be
concerned with the following class of metrics
\begin{equation}
ds^2= \frac{l^2}{r^2}\bigg(-\frac{dt^2}{r^{2\nu}}
+2dtd\xi+dr^2+d\vec{x}^2 \bigg), \label{eq:BSmetric}
\end{equation}
where $\vec{x}$ is a $d$-dimensional vector, $l$ is a constant
associated to the curvature of the space, and $\nu$ is the so-called
dynamical exponent (this is also usually denoted by $z=\nu+1$). For
$\nu=0$, the metric (\ref{eq:BSmetric}) corresponds to anti-de
Sitter spacetime and enjoys the full relativistic conformal
symmetry. For an arbitrary value of the exponent $\nu$, this metric
exhibits as isometries the Galilei transformations on the space
$(t,\vec{x})$ as well as the dilatations transformations. Indeed,
apart from the spacetime translations
$(t,\vec{x})\mapsto(t+b,\vec{x}+\vec{\delta})$ and spatial rotations
in the $\vec{x}$-planes, $\vec{x}\mapsto R\,\vec{x}$ with
$R\in\mathop{\rm SO}(d)$, the metric (\ref{eq:BSmetric}) is
invariant under the Galilean boosts
\begin{equation}\label{eq:boostG}
\varphi_{\vec{v}}(t,\xi,r,\vec{x})=(t,
\xi+\vec{v}\cdot\vec{x}-\frac{1}{2}\vert\vec{v}\vert^2 t,r,
\vec{x}-t\vec{v}),
\end{equation}
as well as under dilations
\begin{equation}\label{eq:dilatations}
\varphi_{a}(t,\xi,r,\vec{x})= (e^{(1+\nu)a}\,t,
e^{(1-\nu)a}\,\xi,e^a\,r,e^a\,\vec{x}).
\end{equation}
For the special value $\nu=1$ of the dynamical exponent, the metric
admits additionally a special conformal transformation given by the
diffeomorphism
\begin{equation}\label{eq:sct}
\varphi_{\kappa}(t,\xi,r,\vec{x})= \bigg(\frac{t}{1-\kappa{t}},
\xi-\frac{\kappa(\vert\vec{x}\vert^2+r^2)}{2(1-\kappa{t})},
\frac{r}{1-\kappa{t}},\frac{\vec{x}}{1-\kappa{t}}\bigg),
\end{equation}
and it is only in this case that the metric is said to enjoy the
full Schr\"odinger symmetry.

The symmetry transformations above are generated for each of its
infinitesimal parameters $\epsilon$ by a Killing vector $K$ in the
standard way, namely
$\varphi_{\epsilon}=\varphi_0+i\epsilon{K}+O(\epsilon^2)$, which
gives rise to the following set of Killing vectors
\begin{eqnarray}
M_{ij} =-i(x_{i}\partial_{j}-x_{j}\partial_{i}),
&&\hspace{-6mm}\qquad
K_{i}=-i(x_{i}\partial_{\xi}-t\partial_{i}), \nonumber\\
P_{i}  =-i\partial_{i},\qquad H&=&-i\partial_{t},\qquad
N=-i\partial_{\xi},\nonumber\\
D      =-i[(1+\nu)t\partial_{t}&&\hspace{-6mm}{}
+(1-\nu)\xi\partial_{\xi}+r\partial_{r}+x^{i}\partial_{i} ].
\end{eqnarray}
For $\nu=1$ (i.e. $z=2$), we find the additional special conformal
generator
\begin{equation}
C=-i\left(t^{2}\partial_{t}
-\frac{|\vec{x}|^2+r^{2}}{2}\partial_{\xi}+tr\partial_{r}
+tx^{i}\partial_{i}\right).
\end{equation}

These Killing vectors realize the algebra whose non-vanishing
commutation relations are
\begin{eqnarray}
[M_{ij},M_{kl}]=i(\delta_{ik}M_{jl}&&\hspace{-6mm}
{}-\delta_{jk}M_{il}
+\delta_{il}M_{kj}-\delta_{jl}M_{ki}),\nonumber\\
{}[M_{ij},P_{k}]=i(\delta_{ik}P_{j}-\delta_{jk}P_{i}&&\hspace{-6mm}),\qquad
[M_{ij},K_{k}]=i(\delta_{ik}K_{j}-\delta_{jk}K_{i}),\hfill~\nonumber\\
{}[P_{i},K_{j}] = -i\delta_{ij}N,\qquad
 &&\hspace{-6mm}[H,K_{i}]=iP_{i},\qquad[D,P_{i}]=iP_{i},\nonumber\\
{}[D,K_{i}]=-i\nu K_{i}\qquad [D,H] &=& i(1+\nu)H,\qquad
[D,N]=i(1-\nu)N.
\end{eqnarray}
Besides, when $\nu=1$,
\begin{equation}
[D,C]=-2iC,\qquad [H,C]=-iD,
\end{equation}
and in this case the algebra corresponds to the Schr\"odinger
algebra.

Notice that the case $\nu=1$ is special not only because it allows
for the additional special conformal generator $C$, but also because
the generator $N$ becomes a central element of the algebra, as it
commutes additionally with the dilation generator $D$ in this case.

In what follows, we will refer to the set of transformations
involving the Galilei transformations $M_{ij}$, $K_i$ $P_i$, $H$,
$N$ and the dilatations $D$ as the \emph{partial\/} Schr\"odinger
symmetry. In the applications to condensed matter physics, each
conformal system turns out to be characterized by the value of $\nu$
its symmetry corresponds to. This helps to identify the candidates
to be the corresponding gravity duals of the form
(\ref{eq:BSmetric}). For example, models describing itinerant
(anti)ferromagnetic materials are thought to be described by the
model with $\nu=2$ (resp.\ with $\nu=1$).

Here, we are mainly interested in the class of metrics admitting the
full (or partial) Schr\"odinger symmetry (\ref{eq:BSmetric}).
Nevertheless, from the gravity viewpoint, it is interesting to
consider first a more general (less symmetric) setting. This is
important to understand what is the appropriate setup these
configurations arise in. With this motivation, we consider an ansatz
of the following form
\begin{equation}\label{eq:ansatz}
d{s}^2=\frac{l^2}{r^2}\left[-F(r)dt^2+2dtd\xi+dr^2+d\vec{x}^2\right],
\end{equation}
where $F$ is the only undetermined structural metric function, and
it only depends on the coordinate $r$. This ansatz corresponds to a
particular class of the so-called Siklos spacetimes
\cite{Siklos:1985}. To be more precise, Siklos spacetimes correspond
to metric (\ref{eq:ansatz}) with a function $F$ that depends on all
the variables except the null coordinate $\xi$. In particular, for a
vanishing structural function $F=0$, we recover the metric of
anti-de~Sitter space in Poincar\'e coordinates, while for $F\ll1$
this metric describes just a perturbation of AdS. In fact, the
metric (\ref{eq:ansatz}) and, more generally, the Siklos spacetimes,
can also be obtained from the AdS one by a generalized Kerr-Schild
transformation (see Appendix). Consequently, they can be though as
describing exact gravitational waves propagating along the AdS
spacetime \cite{Podolsky:1997ik} (AdS-waves). In fact, they become
the particular case, admitting a Killing field, of the more general
exact gravitational waves propagating in presence of a cosmological
constant originally found by Garc\'{\i}a and Pleba\'nski
\cite{Garcia:1981}.\footnote{See
Refs.~\cite{Salazar:1983,Garcia:1983,Ozsvath:1985qn} for further
generalizations.}

In the context of higher-order gravity theories it has been observed
in many cases that the on-shell profile function $F$ of exact
gravitational waves behaves as an exact scalar massive mode
\cite{AyonBeato:2005bm,AyonBeato:2005qq,AyonBeato:2009yq}, since it
satisfies a Klein-Gordon equation
\begin{equation}\label{eq:K-G}
\square{F}=m^2F,
\end{equation}
for some effective mass $m$, and where $\square{}$ stands for the
d'Alambertian operator. For example, in the case of the profile $F$
defining the spacetimes (\ref{eq:BSmetric}), the corresponding mass
is defined in terms of the dynamical exponent as
\begin{equation}\label{eq:mass2(nu)}
m^2=\frac{2\nu(2\nu+D-1)}{l^2},
\end{equation}
and this will allow us to establish a correspondence between many of
the gravity configurations we study and exact scalar massive modes
propagating on these backgrounds.

\section{Square-curvature corrections in arbitrary dimensions\label{sec:D>4}}

The existence of an abundant Schr\"odinger symmetric sector in
higher-dimensional pure gravity becomes clear from the analysis of
Ref.~\cite{Adams:2008zk} where, besides their general arguments, an
explicit example was worked out for a particular five-dimensional
quadratic modification to the Einstein-Hilbert action. Here, we
provide explicitly the general configurations for arbitrary
dimension. In particular, this will lead us to observe the existence
of special critical points in the space of coupling constants at
which logarithmic dependences that necessarily break Schr\"odinger
symmetry appear. Such logarithmic falling-off that emerges at these
special points potentially leads to the definition of weakened
asymptotically AdS boundary conditions. The results of this section
can be thought of as a higher-dimensional generalization of the
recent results of Ref.~\cite{AyonBeato:2009yq} for the
three-dimensional parity-preserving massive gravity introduced in
Ref.~\cite{Bergshoeff:2009hq}.

Unlike the three or four-dimensional cases, in higher dimensions,
three different invariants have to be used to write the most general
quadratic action; namely
\begin{equation}\label{eq:Squad}
S[g_{\mu\nu}]=\int{d}^Dx\sqrt{-g}\left(R-2\lambda+\beta_1{R}^2
+\beta_2{R}_{\alpha\beta}{R}^{\alpha\beta}
+\beta_3{R}_{\alpha\beta\mu\nu}{R}^{\alpha\beta\mu\nu}\right).
\end{equation}
Here, we denote the cosmological constant by $\lambda$ for reasons
that will become clear later. The constants $\beta_i$ are the
coupling constants for the different curvature square modifications.
The field equations obtained by varying the action (\ref{eq:Squad})
with respect to the metric read
\begin{eqnarray}
G_{\mu\nu}+\lambda{g}_{\mu\nu}
+\left(\beta_2+4\beta_3\right)\square{R}_{\mu\nu}
+\frac12\left(4\beta_1+\beta_2\right)g_{\mu\nu}\square{R}
-\left(2\beta_1+\beta_2+2\beta_3\right)\nabla_\mu\nabla_\nu{R}
\nonumber\\\nonumber\\
{}+2\beta_3R_{\mu\gamma\alpha\beta}R_{\nu}^{~\gamma\alpha\beta}
+2\left(\beta_2+2\beta_3\right)R_{\mu\alpha\nu\beta}R^{\alpha\beta}
-4\beta_3R_{\mu\alpha}R_{\nu}^{~\alpha}+2\beta_1RR_{\mu\nu}
\nonumber\\\nonumber\\
{}-\frac12\left(\beta_1{R}^2+\beta_2{R}_{\alpha\beta}{R}^{\alpha\beta}
+\beta_3{R}_{\alpha\beta\gamma\delta}{R}^{\alpha\beta\gamma\delta}
\right)g_{\mu\nu}&=&0.\qquad~\label{eq:squareGrav}
\end{eqnarray}

Before deriving the different class of solutions, we first fix the
cosmological constant $\lambda$ such that the AdS spacetime of
radius $l$ is a solution of the equations (\ref{eq:squareGrav}). By
doing so, we find the following constraint between the cosmological
constant $\lambda$, the AdS radius $l$, and the coupling constants
$\beta_i$
\begin{equation}\label{eq:lambda}
\lambda=-\frac{(D-1)(D-2)}{2l^2}
+\frac{(D-1)(D-4)}{2l^4}\left[(D-1)\left(D\beta_1+\beta_2\right)
+2\beta_3\right].
\end{equation}
>From this, we notice that only in four dimensions the cosmological
constant is related to the AdS radius in the usual way, without
involving the couplings $\beta_i$.

With the choice (\ref{eq:lambda}) for the cosmological constant,
Einstein equations (\ref{eq:squareGrav}) for the AdS waves
(\ref{eq:ansatz}) yield the following equations (see details in the
Appendix)
\begin{eqnarray}
&&\bigg\{(\beta_2+4\beta_3)\left[r^4F''''-2(D-4)r^3F'''\right]\nonumber\\
&&+\big[l^2-2D(D-1)\beta_1+(D-2)(D-8)\beta_2+4(D-2)(D-5)\beta_3\big]
r^2F''
\nonumber\\
&&-(D-2)\left[l^2-2D(D-1)\beta_1-(3D-4)\beta_2-8\beta_3\right]
rF'\bigg\}\frac{\delta_{\mu}^t\delta_{\nu}^t}{2l^2r^2}=0.
\label{eq:SGAdSwave}
\end{eqnarray}

In what follows, we provide a detailed analysis of the equation
(\ref{eq:SGAdSwave}) in order to survey the different solutions
exhaustively.

\subsection{The second order sector\label{subsec:so}}

The first case we will explore is the one for which the fourth-order
differential equation (\ref{eq:SGAdSwave}) reduces to a second-order
equation. This occurs for the special election  $\beta_2=-4\beta_3$,
yielding
\begin{equation}\label{eq:secord}
\frac{1}{l^2}\left[l^2-2D(D-1)\beta_1+12(D-2)\beta_3\right]
\left[r^2F''-(D-2)rF'\right]=0.
\end{equation}
It is clear from this equation that, if additionally one chooses
$\beta_1=[l^2+12(D-2)\beta_3]/[2D(D-1)]$, a full degeneracy appears
and the field equations are satisfied for any profile $F$; this
degenerate class will be analyzed in details in
Subsec.~\ref{subsec:deg}. Then, for
$\beta_1\ne[l^2+12(D-2)\beta_3]/[2D(D-1)]$ the resulting equations
are of second order as in the Lovelock case \cite{Lovelock};
actually, the Lovelock theory which corresponds to
$\beta_2=-4\beta_3$ and $\beta_1=\beta_3$, appears as a particular
case in this analysis. More precisely, for the above inequality (up
to an additive constant that can be removed by coordinate
transformations) the solution is given by
\begin{equation}\label{eq:GR}
F(r)=c_0r^{D-1},
\end{equation}
and coincides with the solution of General Relativity, i.e.\ the one
with all the constants $\beta_i=0$. This solution can be thought of
as the usual General Relativity exact massless scalar mode since the
field equation (\ref{eq:secord}) is proportional to the wave
equation
\begin{equation}\label{eq:massless}
\square{F}=0.
\end{equation}
It is also interesting to note that the solution (\ref{eq:GR})
preserves only the \emph{partial\/} Schr\"odinger symmetry and not
the \emph{full\/} one. In the next subsection, we shall establish
explicitly the existence of a higher order sector enjoying the
\emph{full\/} Schr\"odinger symmetry.

\subsection{The Schr\"odinger invariant sector\label{subsec:Sis}}

For $\beta_2\ne-4\beta_3$, the field equation (\ref{eq:SGAdSwave})
is a fourth-order Euler differential equation. In the generic case,
the space of linearly independent solutions is spanned in power-laws
$F\propto r^\alpha$, where the exponents $\alpha$ are the roots of
the following fourth-degree characteristic polynomial
\begin{equation}\label{eq:charpol}
\alpha(\alpha-D+1) \left[\left(\alpha-\frac{D-1}2\right)^2
-\frac{(D-1)^2}4-\frac{l^2-2(D-1)(D\beta_1+\beta_2)+4(D-4)\beta_3}
{\beta_2+4\beta_3}\right]=0.
\end{equation}
Since the constant solution, i.e.\ $\alpha=0$, can be removed by
coordinate transformations, the general solution is then given by
\begin{subequations}\label{eq:Fgen}
\begin{equation}
F(r)=c_0r^{D-1}+c_+r^{\alpha_+}+c_-r^{\alpha_-},
\end{equation}
where
\begin{equation}\label{eq:alpha_pm}
\alpha_{\pm}=\frac{D-1}2\pm\bigg(\frac{(D-1)^2}4
+\frac{2(D-1)(D\beta_1+\beta_2)-4(D-4)\beta_3-l^2}
{\beta_2+4\beta_3}\bigg)^{1/2},
\end{equation}
\end{subequations}
and where $c_0$ and $c_\pm$ are integrations constants. It is worth
noticing that the solutions (\ref{eq:Fgen}) generate an exact scalar
massive excitation (\ref{eq:K-G}) of mass
\begin{equation}\label{eq:mass_d}
m^2=\frac{2(D-1)(D\beta_1+\beta_2)-4(D-4)\beta_3-l^2}
{l^2(\beta_2+4\beta_3)}.
\end{equation}
To be more precise, the solution (\ref{eq:Fgen}) describes the
superposition of three exact scalar modes given by the massless mode
of General Relativity (\ref{eq:GR}) and two other modes generated by
the squared modifications and both sharing the same mass
(\ref{eq:mass_d}). It is worth pointing out that the solution
(\ref{eq:Fgen}) is valid only when the roots (\ref{eq:alpha_pm}) are
real. This in turn constraints the mass (\ref{eq:mass_d}) to obey
strictly the Breitenlohner-Freedman bound associated to the AdS
space where the waves are propagating on
\cite{Breitenlohner:1982bm,Mezincescu:1984ev}
\begin{equation}\label{eq:BFa}
m^2>-\frac{(D-1)^2}{4l^2}.
\end{equation}

It is also easy to see that taking any pair of the integrations
constants in (\ref{eq:Fgen}) to zero, the isometry group of the
resulting background gets enhanced to the \emph{partial\/}
Schr\"odinger group. The \emph{full\/} Schr\"odinger isometry can
only be achieved by choosing $c_0=0$ and $c_+=0$ while the coupling
constants must be constrained by
\begin{equation}\label{eq:Schro_d}
\beta_2=\frac{D(D-1)}2\beta_1-(3D-2)\beta_3-\frac{l^2}4,
\end{equation}
and this value corresponds to a mass (\ref{eq:mass_d}) given by (see
also (\ref{eq:mass2(nu)}) for $\nu=1$)
\begin{equation}\label{eq:m2d}
m^2=\frac{2(D+1)}{l^2}.
\end{equation}

\subsection{The Logarithmic sectors}

We now turn to the cases for which the roots of the characteristic
polynomial (\ref{eq:charpol}) may have some multiplicities. As it is
well-known, for multiple roots the power laws fail to span all the
linearly independent solutions to Eq.~(\ref{eq:SGAdSwave}) and
additional behaviors exhibiting logarithmic dependence typically
occur. The existence of such \emph{exact\/} logarithmic behaviors
has been established for TMG in
Refs.~\cite{AyonBeato:2004fq,AyonBeato:2005qq}. Moreover, some of
the logarithmic configurations of TMG have been shown to be
compatibles with some relaxed AdS asymptotic
\cite{Grumiller:2008qz,Henneaux:2009pw} and define a sector of the
theory currently known as the Log Gravity sector
\cite{Maloney:2009ck}; see also \cite{GKP}. The relevance of this
sector is that, in three dimensions, it may be holographically dual
at the quantum level to a Logarithmic CFT
\cite{Skenderis:2009nt,Skenderis:2009kd,Grumiller:2009mw,Grumiller:2009sn,Grumiller:2010rm,Alishahiha:2010bw}.
We have shown in \cite{AyonBeato:2009yq} that a similar
\emph{exact\/} Log sector exists also within the context of
NMG.\footnote{See Refs.~\cite{Liu:2009kc,Liu:2009ph} for the
linearized case and Refs.~\cite{Afshar:2009rg,Alishahiha:2010iq} for
the case of other theories as bi-gravity and Born-Infeld gravity
\cite{Gullu:2010pc}. Log gravity was also studied in higher
dimensions recently \cite{Bergshoeff}.}

The first source of multiplicity appears when the roots
(\ref{eq:alpha_pm}) become one single root, which occurs for
\begin{equation}\label{eq:mul_221}
\beta_2=\frac{4\left[l^2-2D(D-1)\beta_1-(D^2-6D+17)\beta_3\right]}
{(D+7)(D-1)},
\end{equation}
and the related double multiplicity solution turns to be
\begin{equation}\label{eq:Fmul_221}
F(r)=c_0r^{D-1}+r^{\frac{D-1}2}\left(c_1\ln{r}+c_2\right).
\end{equation}
For $c_1=0$, two residual sectors having the \emph{partial\/}
Schr\"odinger symmetry still remain, taking either $c_0=0$ or
$c_2=0$. In contrast, the \emph{full\/} Schr\"odinger symmetry is
forbidden for all sectors of the solution.

The global interpretation of the solution (\ref{eq:Fmul_221}) is
also of interest. The configuration represents the superposition of
the massless mode of GR plus two additional exact scalar modes
saturating the BF bound of the AdS space where the waves are
propagating on \cite{Breitenlohner:1982bm,Mezincescu:1984ev}, since
the related profile satisfies
\begin{equation}\label{eq:BF}
\Box{F}=m_{\rm BF}^2F, \qquad m_{\rm
BF}^2\equiv-\frac{(D-1)^2}{4l^2}.
\end{equation}
Notice that this value for the mass is included in the range
(\ref{eq:mass_d}) with $\beta_2$ given by (\ref{eq:mul_221}).

The other multiplicities may appear when one of the two generic
roots (\ref{eq:alpha_pm}) either vanishes or takes the value $D-1$.
In fact, these two possibilities occur simultaneously and in this
case the coupling constants must be restricted as follows
\begin{equation}\label{eq:sdm}
\beta_2=\frac{l^2-2D(D-1)\beta_1+4(D-4)\beta_3}{2(D-1)}.
\end{equation}
The solution with simultaneous double multiplicity is then of the
form
\begin{equation}\label{eq:Fsdm}
F(r)=c_0r^{D-1}+\left(c_1r^{D-1}+c_2\right)\ln{r}.
\end{equation}
At this point of the space of parameters there is no sector
compatible with the \emph{partial\/} Schr\"odinger invariance except
the trivial mode of General Relativity, i.e.\ $c_1=c_2=0$. In
addition of being incompatible with the Schr\"odinger symmetry, the
modes generated by these higher-curvature modifications can not be
interpreted as exact scalar modes as they do not satisfy the
Klein-Gordon equation. However, interesting enough, the profile
(\ref{eq:Fsdm}) can be understood as a ``local'' superposition
\begin{equation}\label{eq:loc_superp}
F = r^{\frac{D-1}2}\bigg[\frac1{r^{\frac{D-1}2}}F|_{c_2=0}\bigg] +
\frac1{r^{\frac{D-1}2}}\bigg[{r^{\frac{D-1}2}}F|_{c_0=0,c_1=0}\bigg],
\end{equation}
of exact massive scalar modes saturating the BF bound [those between
brackets].

\subsection{Below the Breitenlohner-Freedman
bound\label{subsec:deg}}

As it was mentioned in Subsec.~\ref{subsec:Sis} the generic solution
(\ref{eq:Fgen}) is only valid for mass values above the
Breitenlohner-Freedman bound
\cite{Breitenlohner:1982bm,Mezincescu:1984ev}, which constraints the
scalar modes propagating on AdS. However, the modes we consider here
are of gravitational nature and there are {it a priori} no  reasons
for which they must obey the celebrated BF bound. Hence, the sector
having a mass below the Breitenlohner-Freedman bound (\ref{eq:BF}),
\begin{equation}\label{eq:BFb}
m^2<m_{\rm BF}^2,
\end{equation}
might be considered as well, as it has been done in
\cite{Moroz:2009kv} in the context of non-relativistic holographic
correspondence. In this case, the roots (\ref{eq:alpha_pm}) take
complex conjugate values and the solution acquires the following
oscillatory behavior
\begin{equation}\label{eq:FbBF}
F(r)=c_0r^{D-1}+r^{\frac{D-1}2}\left[ c_1\sin\left(l\sqrt{m_{\rm
BF}^2-m^2} \ln{r}\right)+c_2\cos\left(l\sqrt{m_{\rm BF}^2-m^2}
\ln{r}\right)\right],
\end{equation}
where $m^2$ is again given by Eq.~(\ref{eq:mass_d}).

\subsection{The degenerate sector \label{subsec:dege}}

Now, let us discuss the degeneracy in space of solutions and its
relation with the non-renormalization of the dynamical exponent
$\nu$. As it was previously mentioned, it is remarkable that when
the coupling constants are tied in the following manner
\begin{equation}\label{eq:degene}
\beta_1=\frac{l^2+12(D-2)\beta_3}{2D(D-1)}, \qquad
\beta_2=-4\beta_3,
\end{equation}
the metrics (\ref{eq:ansatz}) solves the equations of motion
(\ref{eq:squareGrav}) for any wave profile $F(r)$. In particular,
solutions with \emph{full\/} Schr\"odinger symmetry are admitted in
this case. The specific theories allowing this kind of  degeneracy
are described by the following Lagrangian
\begin{equation}\label{eq:L_d}
R-2\lambda +\beta_3{\cal
L}_{\mathrm{GB}}+\frac{l^2-2(D-3)(D-4)\beta_3} {2D(D-1)}R^2,
\end{equation}
where ${\cal L}_{\mathrm{GB}}=R^2-4R_{\alpha\beta}{R}^{\alpha\beta}
+{R}_{\alpha\beta\mu\nu}{R}^{\alpha\beta\mu\nu}$ is the usual
quadratic Gauss-Bonnet Lagrangian, and the cosmological constant is
fixed as
\begin{equation}
\lambda=-\frac{(D-1)}{4l^2}
\left(D-\frac{4(D-3)(D-4)}{l^2}\beta_3\right). \label{eq:lambda_deg}
\end{equation}
There are two interesting cases included within this family of
degenerate theories. The simplest one is achieved for $\beta_3=0$
and is described by
\begin{equation}\label{eq:L_ds}
R-2\lambda-\frac1{8\lambda}R^2.
\end{equation}
Such fine-tuning in the coupling constant of the quadratic term is
crucial in order to obtain \emph{full\/} Schr\"odinger invariant
configurations and more general profiles; for any other coupling
constant the theory belongs to the ones considered in
Subsec.~\ref{subsec:so} and the \emph{full\/} Schr\"odinger
invariance is forbidden. It is interesting to note that the
Lagrangian (\ref{eq:L_ds}) is the precise combination that allows
the existence of Lifshitz black holes
\cite{Cai:2009ac,AyonBeato:2010tm} for the gravity theory with
$R^2-$corrections. It is also remarkable that for such a fine-tuning
of the coefficients, the Lagrangian, which adopts the $f(R)$-form,
can not be reduced to a scalar-tensor theory through the standard
frame-changing trick \cite{AyonBeato:2010tm}.

Another interesting case occurs for
\[
\beta_3=\frac{l^2}{2(D-3)(D-4)},
\]
and in this case the Lagrangian becomes
\begin{equation}\label{eq:L_iCS}
R-2\lambda -\frac{(D-1)(D-2)}{8(D-3)(D-4)\lambda}{\cal
L}_{\mathrm{GB}}.
\end{equation}
This last case includes the point of the space of parameters where
the Einstein-Gauss-Bonnet gravity coincides with the Chern-Simons
gravity in $D=5$ \cite{Zanelli}, that is
$\beta_1=-\beta_2/4=\beta_3=-3/(4\lambda)$. At this point, the
theory exhibits local gauge invariance under the AdS group
SO($4,2$). A particular feature of Chern-Simons (CS) gravity is that
its space of solutions is highly degenerate, and thus it is not
necessarily surprising that all the metrics (\ref{eq:ansatz}) solve
the equations of motion.\footnote{In fact, a similar behavior occurs
for statics configurations \cite{Dotti:2006cp}.} It is worth
pointing out that the degeneracy that appears at
$\beta_1=-\beta_2/4=\beta_3=-3/(4\lambda)$ in $D=5$ is exactly what
happens in the case $c=0$ and $a\Lambda=3/4$ of
Ref.~\cite{Adams:2008zk} (see Eq.~(2.25) therein), where degenerate
solutions arise.

We have confirmed by explicit calculation that a similar feature is
found in arbitrary number of dimensions as long as one choose the
coupling constants for the Lovelock Lagrangian to exhibit local
gauge invariance. More precisely, if one considers the theory
defined by the action
\begin{equation}\label{eq:hhhhh}
\int d^D x \sqrt{-{g}}\ \sum\limits_{n=0}^{\left[\frac{D-1}2\right]}
\frac{\beta_{n}}{2^{n}n!} \delta _{\lbrack \sigma _{1}}^{\mu
_{1}}\delta _{\rho _{1}}^{\nu _{1}}\ldots\delta _{\sigma _{n}}^{\mu
_{n}}\delta _{\rho _{n}]}^{\nu _{n}}
\prod\limits_{r=1}^{n}{R}_{\quad~~\mu_{r}\nu_{r}}^{\sigma _{r}\rho
_{r}},
\end{equation}
there always exists a special choice of coupling constants
$\beta_1$, $\beta_2$, $\beta_3$, \ldots, such that the above
Lagrangian can be written as a Chern-Simons form in odd dimensions
\cite{Zanelli} and as a Pfaffian form in even dimensions. For the
theories defined with such fine tuning, it turns out that the metric
(\ref{eq:BSmetric}) is admitted as solution for arbitrary $\nu $. In
the language of \cite{Adams:2008zk} this would mean that, if such a
precise tuning between coupling constants $\beta_i$ is considered,
the dynamical exponent $z=\nu+1$ does not get renormalized. Here, we
point out that this non-renormalization is associated to the
enhancement of local (AdS) symmetry at the \emph{Chern-Simons
point}. The question remains whether all the cases (\ref{eq:degene})
exhibit enhancement of symmetry that permits to explain the
non-renormalization of the dynamical exponent in a natural way. To
answer this question one has to be reminded of the fact that the
enhancement of symmetry that occurs at the Chern-Simons point may
also explain the existence of other degenerate points in the moduli
space. That is, the Schr\"odinger invariant backgrounds are such
that changes in the values of $\beta_i$ may be absorbed in a
redefinition of the parameters. In turn, the family of theories that
admit solutions with arbitrary $\nu$ would be parameterized by those
changes of the couplings $\beta_i$ that induce renormalization of
$l$ leaving $\nu$ unchanged.

It is worth mentioning that the AdS wave-like solutions we found
here also appear in other theories with higher-curvature (and not
only square-curvature) terms. A particular case is the family of
theories considered in \cite{Julio}. In the next section we will
consider another type of theories which consists of non-polynomial
corrections of the invariants.

\section{Non-polynomial corrections\label{sec:mggt}}

In this section we will extend our previous results by considering a
gravity theory including modifications that are more general than
the quadratic curvature terms discussed above. In fact, we will
consider the most general action depending on the three curvature
invariants $R$, ${R}_{\alpha\beta}{R}^{\alpha\beta}$ and
${R}_{\alpha\beta\mu\nu}{R}^{\alpha\beta\mu\nu}$, namely\footnote{It
can be seen that similar results are obtained if one allows for more
general modifications, like Lagrangians with functions $f(X)$ of
other invariants like $X=R_{\alpha\mu\beta\nu} R^{\alpha\beta}
R^{\mu\nu}$.}
\begin{equation}\label{eq:Sgener}
S[g_{\mu\nu}]=\int{d}^Dx\sqrt{-g}\left[R-2\tilde{\lambda}
+f\left(R,{R}_{\alpha\beta}{R}^{\alpha\beta},
{R}_{\alpha\beta\mu\nu}{R}^{\alpha\beta\mu\nu}\right) \right],
\end{equation}
where $f$ is a smooth function of the three quadratic curvature
invariants.

Now, let us denote the cosmological constant by $\tilde{\lambda}$ to
emphasize the difference with the previous case (\ref{eq:Squad}).
The field equations obtained by varying the action (\ref{eq:Sgener})
give rise to fourth order equations\footnote{It is also possible to
consider gravity actions depending on invariants constructed with
orther-$k$th derivatives of the curvature. In this case the
resulting equations would be of order $2(k+2)$.}
\begin{eqnarray}
G_{\mu\nu}+\tilde{\lambda}g_{\mu\nu}-\frac{f}2g_{\mu\nu}
+\left(g_{\mu\nu}\square-\nabla_\mu\nabla_\nu+R_{\mu\nu}\right)f_1
+\square\left(f_2{R}_{\mu\nu}\right)
+2f_2R_{\mu\alpha\nu\beta}R^{\alpha\beta}\nonumber\\\nonumber\\
{}+\frac12g_{\mu\nu}\nabla_\alpha\left(2R^{\alpha\beta}\nabla_\beta{f_2}
+f_2\nabla^\alpha{R}\right)
-\nabla_{(\mu}\left(2R_{\nu)}^{~~\alpha}\nabla_\alpha{f_2}
+f_2\nabla_{\nu)}{R}\right)\nonumber\\\nonumber\\
{}+2f_3\left(2\square{R}_{\mu\nu}-\nabla_\mu\nabla_\nu{R}
+R_{\mu\gamma\alpha\beta}R_{\nu}^{~\gamma\alpha\beta}\right.
+\left.2R_{\mu\alpha\nu\beta}R^{\alpha\beta}
-2R_{\mu\alpha}R_{\nu}^{~\alpha}\right)
\nonumber\\\nonumber\\
{}+4\left(R_{\mu\alpha\nu\beta}\nabla^\beta
+2\nabla_\alpha{R}_{\mu\nu}
-2\nabla_{(\mu}R_{\nu)\alpha}\right)\nabla^\alpha{f_3}&=&0,\qquad~
\label{eq:generGrav}
\end{eqnarray}
where
\begin{equation}\label{eq:f1,f2,f3}
f_1\equiv\frac{\partial{f}}{\partial{R}},\qquad
f_2\equiv\frac{\partial{f}}{\partial{R}_{\alpha\beta}{R}^{\alpha\beta}},\qquad
f_3\equiv\frac{\partial{f}}{\partial{R}_{\alpha\beta\mu\nu}
{R}^{\alpha\beta\mu\nu}}.
\end{equation}
We will now argue that in the particular case of AdS-waves
(\ref{eq:ansatz}) the previous equations can be translated into the
equations arising from a square-curvature modification
(\ref{eq:squareGrav}) as studied in the previous section. Notice
that the curvature invariants of an AdS-wave in any dimension are
independent of the specific profile function $F$ and thus are
specified by the AdS space constant invariants
\begin{equation}\label{eq:invariantsD}
{R}=-\frac{D(D-1)}{l^2},\qquad
{R}_{\alpha\beta}{R}^{\alpha\beta}=\frac{D(D-1)^2}{l^4},\qquad
{R}_{\alpha\beta\mu\nu}{R}^{\alpha\beta\mu\nu}=\frac{2D(D-1)}{l^4}.
\end{equation}
This in turn implies that both the function $f$ and its derivatives
$f_i$, $i=1,2,3,$ evaluated on AdS-waves solutions (conformally
pp-waves) become constants denoted by
\begin{equation}\label{eq:fconst}
\tilde{f}\equiv\left.f\right|_{ds^2_\mathrm{AdS}},\qquad
\tilde{f}_i\equiv\left.f_i\right|_{ds^2_\mathrm{AdS}}.
\end{equation}
These two properties imply that the resulting equations coincide
exactly with those of the square-modified gravity
(\ref{eq:squareGrav}) with the following identifications
\begin{equation}\label{eq:beta2f}
\beta_1=\frac{\tilde{f}_1}{2R}=-\frac{l^2\tilde{f}_1}{2D(D-1)},\qquad
\beta_2=\tilde{f}_2,\qquad\beta_3=\tilde{f}_3,
\end{equation}
while the cosmological constant is given by
\begin{equation}\label{eq:cosm_const}
\lambda=\tilde{\lambda}-\frac{\tilde{f}}2-\frac{D(D-1)}{4l^2}
\left\{\tilde{f}_1
-\frac2{l^2}\left[(D-1)\tilde{f}_2+2\tilde{f}_3\right]\right\}.
\end{equation}
The relations above establish a correspondence between the
configurations of square-modified gravity and those considered in
(\ref{eq:Sgener}). Starting from this observation, we can easily
summarize some properties relative to this generic theory of
modified gravity. The first observation is that an AdS space of
radius $l$ is a vacuum configuration of the generalized modified
gravity (\ref{eq:Sgener}) if the cosmological constant is
constrained to be
\begin{equation}\label{eq:lambdag}
\tilde{\lambda}=-\frac{(D-1)(D-2)}{2l^2}
+\frac{\tilde{f}}2+\frac{(D-1)}{l^2}\left\{\tilde{f}_1
-\frac2{l^2}\left[(D-1)\tilde{f}_2 +2\tilde{f}_3\right]\right\}.
\end{equation}
Being the cosmological constant fixed by (\ref{eq:lambdag}), we can
continue our analysis following the lines of the previous section.
For example, the second-order sector of this theory is found for
\begin{equation}\label{eq:secordg}
\tilde{f}_1\ne-1-\frac{12(D-2)\tilde{f}_3}{l^2},\qquad
\tilde{f}_2=-4\tilde{f}_3,
\end{equation}
while the full degeneracy is obtained for
\begin{equation}\label{eq:degeneg}
\tilde{f}_1=-1-\frac{12(D-2)\tilde{f}_3}{l^2},\qquad
\tilde{f}_2=-4\tilde{f}_3.
\end{equation}
In contrast, if $\tilde{f}_2\ne-4\tilde{f}_3$, the field equations
are of fourth order, and the generic configurations turn out to be
the mode superposition (\ref{eq:Fgen}), where the roots defining the
power-law massive modes are now given by
\begin{eqnarray}
\alpha_{\pm}=\frac{D-1}2\pm\bigg(\frac{(D-1)^2}4
\label{eq:alpha_pmg}+\frac{-l^2(1+\tilde{f}_1)+2(D-1)\tilde{f}_2
-4(D-4)\tilde{f}_3}{\tilde{f}_2+4\tilde{f}_3}\bigg)^{1/2},
\end{eqnarray}
and the mass associated to the scalar excitation $F$ reads
\begin{equation}\label{eq:mass_dg}
m^2=\frac{-l^2(1+\tilde{f}_1)+2(D-1)\tilde{f}_2-4(D-4)\tilde{f}_3}
{l^2(\tilde{f}_2+4\tilde{f}_3)}.
\end{equation}
These configurations contain sub-sectors which have \emph{partial\/}
Schr\"odinger symmetry, exhibiting the \emph{full\/} symmetry only
for
\begin{equation}\label{eq:Schro_dg}
\tilde{f}_2=-\frac{l^2}4(1+\tilde{f}_1)-(3D-2)\tilde{f}_3,
\end{equation}
with the corresponding mass (\ref{eq:m2d}) in this case.

The logarithmic sector which includes the modes saturating the BF
bound (\ref{eq:Fmul_221}) appears for a generic gravity modification
satisfying
\begin{equation}\label{eq:mul_221g}
\tilde{f}_2=\frac{l^2(1+\tilde{f}_1)-(D^2-6D+17)\tilde{f}_3}
{(D+7)(D-1)}.
\end{equation}
The other Log sector allowing a ``local'' superposition of modes
saturating the BF bound [see Eqs.~(\ref{eq:Fsdm}) and
(\ref{eq:loc_superp})] is possible if
\begin{equation}\label{eq:sdmg}
\tilde{f}_2=\frac{l^2(1+\tilde{f}_1)+4(D-4)\tilde{f}_3}{2(D-1)}.
\end{equation}

\section{Conformally invariant corrections \label{sec:Conf_corr}}

Even though the existence of Schr\"odinger invariant solutions is
not particularly attached to the conformal invariance of the
higher-curvature corrections, we find interesting to investigate the
particular case of the Einstein gravity supplemented by the
conformal Weyl Lagrangian in $D$-dimensions; namely
\begin{equation}\label{eq:BWE}
S[g_{\mu\nu}]=\int{d}^Dx\sqrt{-g}\left[R-2\Lambda
+\frac{1}{2\alpha_w}\left(C_{\alpha\beta\mu\nu}
C^{\alpha\beta\mu\nu}\right)^{D/4}\right],
\end{equation}
where $\alpha_w$ is a coupling constant and $C_{\alpha\beta\mu\nu}$
is the Weyl tensor, whose quadratic contraction reads
\begin{equation}
C_{\alpha\beta\mu\nu} C^{\alpha\beta\mu\nu}=\frac2{(D-1)(D-2)}R^2
-\frac4{D-2}R_{\alpha\beta}R^{\alpha\beta}
+R_{\alpha\beta\mu\nu}R^{\alpha\beta\mu\nu}.\label{eq:W^2}
\end{equation}
It is easy to verify that the term in the action (\ref{eq:BWE}) that
involves the contraction of the quadratic Weyl tensor is conformally
invariant.

Using the notation introduced in the previous section [see
Eq.~(\ref{eq:f1,f2,f3})], we obtain
\begin{eqnarray}
f_1&=&\frac{DR}{2(D-1)(D-2)\alpha_w}\left(C_{\alpha\beta\mu\nu}
C^{\alpha\beta\mu\nu}\right)^{(D-4)/4},\nonumber\\
f_2&=&-\frac{D}{2(D-2)\alpha_w}\left(C_{\alpha\beta\mu\nu}
C^{\alpha\beta\mu\nu}\right)^{(D-4)/4},\label{eq:f1c,f2c,f3c}\\
f_3&=&\frac{D}{8\alpha_w}\left(C_{\alpha\beta\mu\nu}
C^{\alpha\beta\mu\nu}\right)^{(D-4)/4}\nonumber.
\end{eqnarray}
For dimensions $D>4$, the above scalars $f_i$ are proportional to a
positive power of the square of the Weyl tensor. The related
constants $\tilde{f}$ and $\tilde{f}_i$, being obtained by
evaluating the scalars in their AdS values (\ref{eq:fconst}), vanish
since AdS space is conformally flat and hence the Weyl tensor
vanishes identically. This means that, in what regards to the
AdS-wave solutions, the conformally invariant modification of
Einstein gravity gives no correction for dimensions $D>4$. This case
is described by the branch (\ref{eq:secordg}) and the resulting
configuration is the one of General Relativity (\ref{eq:GR}) which
yields only the \emph{partial\/} Schr\"odinger symmetry. In four
dimensions the situation is quite different since the resulting
theory (\ref{eq:BWE})-(\ref{eq:W^2}) corresponds to the
square-curvature corrections that we have considered in Sec.
\ref{sec:D>4} with
$$
\beta_1=\frac{1}{6\alpha_w},\qquad \beta_2=-\frac{1}{\alpha_w},
\qquad \beta_3=\frac{1}{2\alpha_w}.
$$
As a consequence, except for the second-order and degenerate sectors
which require $\beta_2=-4\beta_3$, all the other sectors of
solutions described in Sec.~\ref{sec:D>4} are present for this
four-dimensional Einstein-Weyl gravity theory.

\section{Parity-violating Chern-Simons modification \label{sec:C-Smod}}

Now, let us move to study another interesting example of
gravitational model in four dimensions. This is the Jackiw-Pi theory
\cite{Jackiw:2003pm}, usually referred to as the Chern-Simons
modified gravity in four dimensions. This model is defined by
supplementing the Einstein-Hilbert action with a different (parity
violating) quadratic term in the curvature, yielding the total
action
\begin{equation}\label{eq:JP}
\hat{S}[{g}_{\mu\nu}]=\int{d}^4x\sqrt{-{g}}\left({R}-2\Lambda
+\frac{\theta}{4}\,^{\ast}{R}_{\alpha\beta\mu\nu}
{R}^{\alpha\beta\mu\nu}\right).
\end{equation}
Here $\theta$ is a local Lagrange multiplier that couples to the
Pontryagin density
$^{\ast}{R}_{\alpha\beta\mu\nu}{R}^{\alpha\beta\mu\nu}$, constructed
via the dual curvature tensor
$^{\ast}{R}^{\alpha~\mu\nu}_{~\beta}=\frac{1}{2}
{\eta}^{\rho\sigma\mu\nu}{R}^{\alpha}_{~\beta\rho\sigma}$, where
${\eta}_{\rho\sigma\mu\nu}$ is the volume 4-form.\footnote{Here
$\eta_{t{\xi}rx}=\sqrt{-g}$ ($\eta^{t{\xi}rx}=-1/\sqrt{-g}$).} The
coupling is such that $\theta$ has dimensions of [length]$^{2}$.
Note that the action (\ref{eq:JP}) has to be distinguished from the
Chern-Simons gravitational theories of Ref.~\cite{Zanelli}, which
exist in odd dimensions and correspond to a particular case of
Lovelock Lagrangian \cite{Lovelock}.

The inclusion of the non-dynamical field $\theta $ comes from the
fact that the Pontryagin form
$^{\ast}{R}_{\alpha\beta\mu\nu}{R}^{\alpha\beta\mu\nu}$ is a total
derivative. As a consequence, the variation with respect to the
metric brings an additional piece to the Einstein equations, which
corresponds to a four-dimensional version of the Cotton tensor whose
definition obviously depends on $\theta$. Since the field $\theta$
can be fixed arbitrarily, the diffeomorphism invariance is broken.
Nevertheless, the conservation of the equations of motion makes
diffeomorphism symmetry to be restored dynamically. That is, the
consistency of the theory imposes that only geometries with
vanishing Pontryagin form are allowed as solutions. The same
conclusion is obtained by considering the non-dynamical field
$\theta$ as a Lagrange multiplier; see Ref.~\cite{Jackiw:2007br} for
a careful digression on this point.

The equations of motion derived from the Jackiw-Pi action
(\ref{eq:JP}) then take the form
\begin{equation}\label{eq:CSeom}
{C}_{\mu\nu}+{G}_{\mu\nu}-\frac{3}{l^{2}}{g}_{\mu\nu}=0,
\end{equation}
where, as mentioned above, ${C}_{\mu\nu}$ is a sort of
generalization of the three-dimensional Cotton tensor that appears
in the equations of motion of TMG, given by\footnote{There are other
generalizations of the Cotton tensor in higher dimensions, see e.g.\
Ref.~\cite{Garcia:2003bw}.}
\begin{equation}\label{eq:Cotton}
{C}^{\mu\nu}={\nabla}_{\alpha}\left( {\nabla}_{\beta}\theta\,
^{\ast}{R}^{\alpha(\mu|\beta|\nu)}\right).
\end{equation}

The conservation of the Einstein equations yields the additional
constraint
\begin{equation}\label{eq:nuevovinculo}
{\nabla}^{\mu}{C}_{\mu\nu}=\frac{1}{8}
\,^{\ast}{R}_{\alpha\beta\rho\sigma}
{R}^{\alpha\beta\rho\sigma}{\nabla}_{\nu}\theta=0,
\end{equation}
which imposes that all solutions of the Jackiw-Pi theory
(\ref{eq:JP}) must obey
$^{\ast}{R}_{\mu\nu\rho\sigma}{R}^{\mu\nu\rho\sigma}=0$. It is worth
noticing that all Siklos spacetimes, being AdS waves, satisfy this
necessary condition as they have vanishing Pontryagin invariant. In
fact, we have $^{\ast } {C}_{\mu \nu \rho \sigma }{C}^{\mu \nu \rho
\sigma }=0$ which in turn implies that  $^{\ast}{R}_{\mu \nu \rho
\sigma }{R}^{\mu \nu \rho \sigma }=0$ since both invariants agree
(for a detailed analysis, see the first Appendix of
Ref.~\cite{Grumiller:2007rv}). This is a promising scenario for the
search of Schr\"odinger symmetric configurations. Indeed, we will
now show  that the Jackiw-Pi theory (\ref{eq:JP}) admits solutions
with Schr\"odinger symmetry. In order to realize this task, the
first step is to make an ``educated'' choice for the breaking of
diffeomorphism invariance by considering\footnote{It is possible to
show that there exists an infinite family of elections compatible
with the existence of Schr\"odinger symmetry, and we are just
presenting the simplest one.}
\begin{equation}\label{eq:theta}
\theta=\frac1{\alpha_p}\frac{x}r,
\end{equation}
where $\alpha_p$ is a coupling constant with dimensions of
[length]$^{-2}$. This choice reduces the variational Cotton-like
tensor (\ref{eq:Cotton}) to the single expression
\begin{equation}\label{eq:CottonSiklos}
{C}_{\mu\nu}=-\frac{r^2}{2\alpha_pl^2} \left(\frac{F''}r\right)'
\delta_{\mu}^t\delta_{\nu}^t.
\end{equation}
Taking all this into account, the Jackiw-Pi equations
(\ref{eq:CSeom}) become
\begin{equation}\label{eq:JPtt}
-\frac{r^2}{2\alpha_pl^2}\left[\frac{1}{r^{1-\alpha_pl^2}}
\left(\frac{F'}{r^{\alpha_pl^2}}\right)'\right]'
\delta_{\mu}^t\delta_{\nu}^t=0.
\end{equation}
For generic values of the coupling constant $\alpha_p$, with
$\alpha_p\ne-1/l^2$ and $\alpha_p\ne2/l^2$, the solution of this
equation is given by
\begin{equation}\label{eq:apg}
F(r)=c_0r^3+c_1r^{1+\alpha_pl^2},
\end{equation}
where we have discarded the additive constant removable through a
coordinate transformation. On the other hand, for the critical
values $\alpha_p=-1/l^2$ and $\alpha_p=2/l^2$, we obtain the
following logarithmic branches
\begin{eqnarray}
F(r) &=& c_0r^3+c_1\ln{r},\label{eq:apl1}\\
F(r) &=& c_0r^3+c_1r^3\ln{r}, \label{eq:apl-2}
\end{eqnarray}
respectively. Solutions (\ref{eq:apl1}) and (\ref{eq:apl-2}) enjoy
the partial Schr\"odinger symmetry for $c_1=0$ while the generic
solution (\ref{eq:apg}) is also partially Schr\"odinger invariant as
long as one of the two constants is set to zero. The analogy with
the theories studied previously is complete since the generic
solution (\ref{eq:apg}) have the \emph{full\/} Schr\"odinger
symmetry for the special fine tuning
$$
\alpha_p=-\frac3{l^2}.
$$
There also exists a close relation to TMG: Identifying the
three-dimensional topological mass in term of the four-dimensional
coupling constant as $\mu=\alpha_pl$,\footnote{Notice we are using
here the definition of the topological mass for example of
Ref.~\cite{Henneaux:2009pw}} both the generic solution
(\ref{eq:apg}) and the critical one (\ref{eq:apl1}) can be
represented for $c_0=0$ as a warped product having as base the
corresponding TMG configurations of
Refs.~\cite{AyonBeato:2004fq,Dereli:2000fm,Olmez:2005by,%
AyonBeato:2005qq,Carlip:2008eq,Gibbons:2008vi} with a real line
fiber generated by the spatial direction along the coordinate $x$;
namely
\begin{equation}\label{eq:rel2TMG}
d{s}^2_{\rm{JP}}=d{s}^2_{\rm{TMG}}+\frac{l^2}{r^2}dx^2.
\end{equation}

The other critical solution (\ref{eq:apl-2}) which corresponds to
$\alpha_p=2/l^2$, does not allow such a simple representation in
term of the remaining three-dimensional critical TMG solution with
topological mass $\mu=1/l$. However, since this three-dimensional
case turns out to be a consistent asymptotically AdS configuration
in TMG despite its weakened logarithmic decay
\cite{Grumiller:2008qz,Henneaux:2009pw}, it would be very
interesting to check if the above critical solution (\ref{eq:apl-2})
for $\alpha_p=2/l^2$ is also an asymptotically AdS solution of the
Jackiw-Pi theory, relaxing the standard asymptotic conditions known
for General Relativity \cite{Henneaux:1985tv}.

\section{Conclusions\label{sec:conclu}}

In this paper, we have been concerned with a special class of Siklos
spacetimes that contains the Schr\"odinger invariant metrics as
particular cases. Our main purpose was to identify pure gravity
theories that exhibit solutions of this special kind. We began by
considering the Einstein gravity augmented by square-curvature
corrections in arbitrary dimensions $D$. In this case, we have shown
that for a particular choice of the coupling constants Schr\"odinger
invariant metrics are allowed as solutions, while other choice
yields to a sector of solutions with logarithmic falling-off. We
have also observed the existence of a degenerate sector whose space
of solutions contains all these particular Siklos spacetimes,
without restricting the profile function $F$ that appears in the
metric. We discussed the relation of this degeneracy to the
non-renormalization of the dynamical exponent $z=\nu+1$ observed at
special points of the moduli space. We further extended our results
to a larger set of gravity theories, whose Lagrangians are given by
arbitrary functions of the square-curvature invariants. This was
achieved by establishing a correspondence with square-curvature
models discussed first. All the sectors studied for the
square-curvature actions were shown to also appear in this more
general class of models. As a special example, we considered the
theory defined by adding to the Einstein-Hilbert action
non-polynomial conformally invariant corrections in arbitrary
dimension $D$. In this case, we observed that only in $D=4$, which
corresponds to a particular case of square-curvature action, this
theory presents genuinely new features, as for the case of higher
dimensions the type of solution we discuss simply reduces to the
solutions of $D$-dimensional General Relativity. Finally, we have
also analyzed a completely different higher-order theory of gravity,
the so-called Chern-Simons modification of four-dimensional General
Relativity proposed by Jackiw and Pi. This model involves a
non-dynamical field that plays the role of a Lagrange multiplier
which forces the Pontryagin density to vanish. We have shown that
for some precise choices of the non-dynamical field, the
parity-violating Chern-Simons modification of General Relativity
exhibits both the Schr\"odinger invariant and the logarithmic
sectors as exact solutions.

In two recent papers, \cite{AyonBeato:2010tm} and
\cite{AyonBeato:2009nh}, we have shown that the Einstein gravity
together with appropriated square-curvature corrections in arbitrary
dimensions $D$ admits black holes configurations that asymptote the
Lifshitz spacetimes \cite{Kachru:2008yh}. A natural continuation of
the work would be finding black holes with Schr\"odinger asymptotic
for (some of) the higher order gravity theories considered in this
paper. It would be interesting to obtain black hole configurations
which asymptote the metrics (\ref{eq:BSmetric}) for arbitrary
$\nu\not=1$ and, consequently, to get black holes with partial
Schr\"odinger asymptotic. This question is of physical relevance as
the black hole solutions that have been derived so far in the
context of string theory and through the null Melvin twist (see
\cite{Adams:2008wt} and \cite{Herzog:2008wg}) possess the full
Schr\"odinger symmetry asymptotically by construction and hence
correspond only to the class of solutions with $\nu=1$. This
prevents us from the possibility to use an holographic description
for the finite temperature effects of condensed matter systems
having $\nu\neq1$.

Finally, let us conclude with few words about some classes of
metrics that depends on time, and therefore the time translation is
not longer an isometry, while the dilatation (\ref{eq:dilatations})
or the special conformal transformations (\ref{eq:sct}) still act as
isometries \cite{Naka}. A particular class of such metrics can be
written in the following form
\begin{equation}\label{eq:BSmetrictime}
ds^2= \frac{l^2}{r^2}\left[-G\left(\frac{t}{r^{1+\nu}}\right)
\frac{dt^2}{r^{2\nu}}+2dtd\xi+dr^2+d\vec{x}^2 \right],
\end{equation}
where $G$ is an arbitrary function of the argument $t/r^{1+\nu}$. In
the sectors of solutions studied in this paper, the solutions depend
on arbitrary constants denoted by $c_i$ or $c_{\pm}$. However, it is
easy to see that these constants can be replaced by arbitrary
functions of the time $t$. As a consequence, all the power law
solutions derived previously of the form $F(r)=c\,r^{\alpha}$ can be
extended to $F(t,r)=c(t)\,r^{\alpha}$ where $c(t)$ is an arbitrary
function of $t$. Therefore, choosing
$c(t)=\tilde{c}/t^{\frac{2\nu+\alpha}{1+\nu}}$, where $\tilde{c}$ is
a constant, makes the solution dilatation invariant for $\nu\not=1$,
while for $\nu=1$ the solution admits the special conformal
transformation as isometry but not the dilatation.

\acknowledgments

Conversations and correspondence with M.\ Aiello, A.\ Garbarz, A.\
Garc\'{\i}a, J.\ Gomis, A.\ Maloney, and J.\ Oliva are acknowledged.
This work has been partially supported by grant 1090368 from
FONDECYT, by grant ACT56 from CONICYT, by grants PICT, PIP and
UBACyT from ANPCyT, CONICET and UBA, and by grant 82443 from
CONACyT.

\appendix

\section{AdS-waves and higher order terms\label{app:hot}}

Let us consider an AdS-wave of the type (\ref{eq:ansatz}). It is
worth noticing that the following formulae is valid even when the
profile has a more general dependence $F=F(u,r,\vec{x})$. We use the
null geodesic vector $k^\mu\partial_\mu=(r/l)\partial_\xi$ that
allows the reinterpretation of these backgrounds as generalized
Kerr-Schild transformations of AdS
\begin{equation}\label{eq:K-S}
g_{\mu\nu}=g^{\mathrm{AdS}}_{\mu\nu}-Fk_{\mu}k_\nu.
\end{equation}
We start with the Ricci tensor, which for a $D$-dimensional AdS-wave
is written as
\begin{equation}\label{eq:Ricci}
R_{\mu\nu}=-\frac{(D-1)}{l^2}g_{\mu\nu}
+\frac12k_{\mu}k_{\nu}\square{F},
\end{equation}
yielding the scalar curvature $R=-D(D-1)/l^2$, exactly the same as
for AdS space. This gives the Einstein tensor
\begin{equation}\label{eq:Einstein}
G_{\mu\nu}=\frac{(D-1)(D-2)}{2l^2}g_{\mu\nu}
+\frac12k_{\mu}k_{\nu}\square{F},
\end{equation}
and the squared-curvature combinations
\begin{eqnarray}
RR_{\mu\nu}&=&\frac{D(D-1)^2}{l^4}g_{\mu\nu}
-\frac{D(D-1)}{2l^2}k_{\mu}k_{\nu}\square{F},\qquad\label{eq:R*Ricci}\\
R_{\mu\alpha}R_{\nu}^{~\alpha}&=&\frac{(D-1)^2}{l^4}g_{\mu\nu}
-\frac{(D-1)}{l^2}k_{\mu}k_{\nu}\square{F}.\label{eq:Ricci*Ricci}
\end{eqnarray}

The following squared-curvature combinations involve explicitly the
Riemann tensor
\begin{eqnarray}
R_{\mu\alpha\nu\beta}R^{\alpha\beta}&=&\frac{(D-1)^2}{l^4}g_{\mu\nu}
-\frac{(D-2)}{2l^2}k_{\mu}k_{\nu}\square{F},\qquad
\label{eq:Riemann*Ricci}\\
R_{\mu\gamma\alpha\beta}R_{\nu}^{~\gamma\alpha\beta}&=&
\frac{2(D-1)}{l^4}g_{\mu\nu}
-\frac{2}{l^2}k_{\mu}k_{\nu}\square{F}.\label{eq:Riemann*Riemann}
\end{eqnarray}

Using the expression for the Ricci tensor (\ref{eq:Ricci}), together
with the null and geodesic properties of $k^\mu$, it is not hard to
verify that
\begin{equation}\label{eq:Box_Ricci}
\square{R}_{\mu\nu}=\frac12k_{\mu}k_{\nu}
\square\left(\square-\frac2{l^2}\right)F.
\end{equation}

If we denote by $K_{\mu\nu}$ the modification to the Einstein
equations (\ref{eq:squareGrav}) coming from the squared curvatures,
the expressions above allow to write this tensor as
\begin{eqnarray}
K_{\mu\nu}&=&-\frac{(D-1)(D-4)}{2l^2}
\left[(D-1)(D\beta_1+\beta_2)+2\beta_3\right]g_{\mu\nu}
\nonumber\\\nonumber\\
&&+\frac12k_{\mu}k_{\nu}\square
\bigg\{(\beta_2+4\beta_3)\square\label{eq:K}-\frac2{l^2}\left[(D-1)(D\beta_1+\beta_2)
-2(D-4)\beta_3\right] \bigg\}F.\nonumber
\end{eqnarray}

Now the Einstein equations (\ref{eq:squareGrav}) become
\begin{eqnarray}
&&\left\{\lambda+\frac{(D-1)(D-2)}{2l^2}-\frac{(D-1)(D-4)}{2l^4}
\left[(D-1)\left(D\beta_1+\beta_2\right)
+2\beta_3\right]\right\}g_{\mu\nu}\nonumber\\\nonumber\\
&&{}+\frac12k_{\mu}k_{\nu}\square
\bigg\{(\beta_2+4\beta_3)\square-\frac1{l^2}\left[2(D-1)(D\beta_1+\beta_2)
-4(D-4)\beta_3-l^2\right]\bigg\}F=0,\nonumber
\end{eqnarray}
from which it follows that the cosmological constant must be chosen
as in Eq.~(\ref{eq:lambda}). Since the d'Alembertian of any function
$\Phi=\Phi(r)$ depending only in the front-wave coordinate $r$
becomes
\begin{equation}\label{eq:box}
\square\Phi=\frac1{l^2}\left[r^2\Phi''-(D-2)r\Phi'\right],
\end{equation}
we obtain finally equation (\ref{eq:SGAdSwave}).


\end{document}